\documentstyle[12pt]{article}

\renewcommand{\bar}[1]{\overline{#1}}

\newcommand{\VEV}[1]{\left\langle{#1}\right\rangle}
\newcommand{\ket}[1]{\,\left|\,{#1}\right\rangle}
\newcommand{\M}{{\cal M}}
\newcommand{\R}{{\cal R}}

\newcommand{\ie}{{\em i.e.}}
\newcommand{\eg}{{\em e.g.}}
\newcommand{\brk}{{\hfill\break}}

\begin{document}
\begin{flushright}
SLAC--PUB--7542\\
June 1997
\end{flushright}
\vfill
\bigskip\bigskip

\thispagestyle{empty}
\flushbottom

\centerline{\Large\bf The Light--Cone Fock State Expansion}
\smallskip 
\centerline{{\Large\bf and Hadron Physics Phenomenology} 
    \footnote{\baselineskip=14pt
     Work supported by the Department of Energy, contract 
     DE--AC03--76SF00515.}}
  
\vfill
  \centerline{\bf Stanley J. Brodsky}
\vspace{8pt}
  \centerline{\em Stanford Linear Accelerator Center}
  \centerline{\em Stanford University, Stanford, California 94309}
  \centerline{e-mail: sjbth@slac.stanford.edu}
\vspace*{0.9cm}
\vfill
\begin{center}
Invited talk at the International Workshop \\
``New Nonperturbative Methods and Quantization on the
Light Cone'' \\
Les Houches, France \\
February 24--March 7, 1997
\end{center}
\vfill
\newpage

\section{Introduction}

The concept of the ``number of constituents" of a relativistic bound
state, such as a hadron in quantum chromodynamics, is not only
frame-dependent, but its value can fluctuate to an arbitrary number of
quanta. Thus when a laser beam crosses a proton at fixed ``light-cone"
time $\tau = t+z/c= x^0 + x^z$, an interacting photon can encounter a
state with any given number of quarks, anti-quarks, and gluons in
flight (as long as $n_q - n_{\bar q} = 3$). The probability amplitude
for each such $n$-particle state of on-mass shell quarks and gluons in
a hadron is given by a light-cone Fock state wavefunction
$\psi_{n/H}(x_i,\vec k_{\perp i},\lambda_i)$, where the constituents
have longitudinal light-cone momentum fractions
\begin{equation}
x_i = \frac{k^+_i}{p^+} = \frac{k^0+k^z_i}{p^0+p^z}\ ,
\quad \sum^n_{i=1} x_i= 1
\ ,
\label{eq:c}
\end{equation}
relative transverse momentum
\begin{equation}
\vec k_{\perp i} \ , \quad \sum^n_{i=1}\vec k_{\perp i} = \vec 0_\perp \ ,
\label{eq:d}
\end{equation}
and helicities $\lambda_i.$ The ensemble \{$\psi_{n/H}$\} of such
light-cone Fock wavefunctions is a key concept for hadronic physics,
providing a conceptual basis for representing physical hadrons (and
also nuclei) in terms of their fundamental quark and gluon degrees of
freedom \cite{BR}.

The light-cone Fock expansion is defined in the following way:  one
first constructs the light-cone time evolution operator $P^-=P^0-P^z$
and the invariant mass operator $H_{LC}= P^- P^+-P^2_\perp $ in
light-cone gauge $A^+=0$ from the QCD Lagrangian. The total
longitudinal momentum $P^+ = P^0 + P^z$ and transverse momenta $\vec
P_\perp$ are conserved, \ie\ are independent of the interactions. The
matrix elements of $H_{LC}$ on the complete orthonormal basis $\{\vert
n >\}$ of the free theory $H^0_{LC} = H_{LC}(g=0)$ can then be
constructed.  The matrix elements $\VEV{n\,|\,H_{LC}\,|\,m}$ connect
Fock states differing by 0, 1, or 2 quark or gluon quanta, and they
include the instantaneous quark and gluon contributions imposed by
eliminating dependent degrees of freedom in light-cone gauge.

In practice it is essential to introduce an ultraviolet regulator in
order to limit the total range of $\VEV{n\,|\,H_{LC}\,|\,m}$, such as
the ``global" cutoff in the invariant mass of the free Fock states:
\begin{equation} \M^2_n = \sum^n_{i=1}
\frac{k^2_\perp + m^2}{x} < \Lambda^2_{\rm global} \ . 
\label{eq:a}
\end{equation} 
One can also introduce a ``local" cutoff to limit the change in
invariant mass $|\M^2_n-\M^2_m| < \Lambda^2_{\rm local}$ which provides
spectator-independent regularization of the sub-divergences associated
with mass and coupling renormalization.

The natural renormalization scheme for the coupling is $\alpha_V(Q)$,
the effective charge defined from the scattering of two
infinitely-heavy quark test charges.  The renormalization scale can
then be determined from the virtuality of the exchanged momentum, as in
the BLM and commensurate scale methods
\cite{BLM,CSR,BrodskyKataevGabaladzeLu}.

In the discretized light-cone method (DLCQ) \cite{DLCQ,Schlad} the
matrix elements $\VEV{n\,|\,H^{\Lambda)}_{LC}\,|\,m}$, are made
discrete in momentum space by imposing periodic or anti-periodic
boundary conditions in $x^-=x^0 - x^z$ and $\vec x_\perp$. Upon
diagonalization of $H_{LC}$, the eigenvalues provide the invariant mass
of the bound states and eigenstates of the continuum.  The projection
of the hadronic eigensolutions on the free Fock basis define the
light-cone wavefunctions. For example, for the proton,
\begin{eqnarray}
\ket p &=& \sum_n \VEV{n\,|\,p}\, \ket n \nonumber \\
&=& \psi^{(\Lambda)}_{3q/p} (x_,\vec k_{\perp i},\lambda_i)\,
\ket{uud} \\[1ex]
&&+ \psi^{(\Lambda)}_{3qg/p}(x_i,\vec k_{\perp i},\lambda_i)\,
\ket{uudg} + \cdots \nonumber
\label{eq:b}
\end{eqnarray}

The light-cone formalism has the remarkable feature that the
$\psi^{(\Lambda)}_{n/H}(x_i, \vec k_{\perp i},\Lambda_c)$ are invariant
under longitudinal boosts; \ie,\ they are independent of the total
momentum $P^+$, $\vec P_\perp$ of the hadron.  Given the
$\psi^{(\Lambda)}_{n/H},$ we can construct any electromagnetic or
electroweak form factor from the diagonal overlap of the LC
wavefunctions\cite{BD}.  Similarly, the matrix elements of the currents
that define quark and gluon structure functions can be computed from
the integrated squares of the LC wavefunctions \cite{BrodskyLepage}.

In general, any hadronic amplitude such as quarkonium decay, heavy
hadron decay, or any hard exclusive hadron process can be constructed
as the convolution of the light-cone Fock state wavefunctions with
quark-gluon matrix elements \cite{BL}
\begin{eqnarray}
\M_{\rm Hadron} &=& \prod_H \sum_n \int
\prod^{n}_{i=1} d^2k_\perp \prod^{n}_{i=1}dx\, \delta
\left(1-\sum^n_{i=1}x_i\right)\, \delta
\left(\sum^n_{i=1} \vec k_{\perp i}\right) \nonumber \\[2ex]
&& \times \psi^{(\Lambda)}_{n/H} (x_i,\vec k_{\perp i},\Lambda_i)\, \M
^{(\Lambda)}_{q,g} \ . 
\label{eq:e}
\end{eqnarray}
Here $\M^{(\Lambda)}_{q,g}$ is the underlying quark-gluon subprocess
scattering amplitude, where the (incident or final) hadrons are
replaced by quarks and gluons with momenta $x_ip^+$, $x_i\vec
p_{\perp}+\vec k_{\perp i}$ and invariant mass above the separation
scale $\M^2_n > \Lambda^2$. The LC ultraviolet regulators thus provide
a factorization scheme for elastic and inelastic scattering, separating
the hard dynamical contributions with invariant mass squared $\M^2 >
\Lambda^2_{\rm global}$ from the soft physics with $\M^2 \le
\Lambda^2_{\rm global}$ which is incorporated in the nonperturbative LC
wavefunctions.  The DGLAP evolution of parton distributions can be
derived by computing the variation of the Fock expansion with respect
to $\Lambda^2_{\rm global}$ \cite {BL}.

The simplest, but most fundamental, characteristic of a hadron in the
light-cone representation, is the hadronic distribution amplitudes
\cite{BL}, defined as the integral over transverse momenta of the
valence (lowest particle number) Fock wavefunction; \eg\ for the pion
\begin{equation}
\phi_\pi (x_i,Q) \equiv \int d^2k_\perp\, \psi^{(Q)}_{q\bar q/\pi}
(x_i, \vec k_{\perp i},\lambda)
\label{eq:f}
\end{equation}
where the global cutoff $\Lambda_{global}$ is identified with the
resolution $Q$.  The distribution amplitude controls leading-twist
exclusive amplitudes at high momentum transfer, and it can be related
to the gauge-invariant Bethe-Salpeter wavefunction at equal light-cone
time $\tau = x^+$.  The $\log Q$ evolution of the hadron distribution
amplitudes $\phi_H (x_i,Q)$ can be derived from the
perturbatively-computable tail of the valence light-cone wavefunction
in the high transverse momentum regime \cite{BL}.

Light-cone quantization methods have had remarkable success in solving
quantum field theories in one-space and one-time dimension---virtually
any (1+1) quantum field theory can be solved using DLCQ. A beautiful
example is ``collinear" QCD:  a variant of $QCD(3+1)$ defined by
dropping all of interaction terms in $H^{QCD}_{LC}$ involving
transverse momenta \cite{Kleb}.  Even though this theory is effectively
two-dimensional, the transversely-polarized degrees of freedom of the
gluon field are retained as two scalar fields.  Antonuccio and Dalley
\cite{AD} have used DLCQ to solve this theory. The diagonalization of
$H_{LC}$ provides not only the complete bound and continuum spectrum of
the collinear theory, including the gluonium states, but it also yields
the complete ensemble of light-cone Fock state wavefunctions needed to
construct quark and gluon structure functions for each bound state. 
Although the collinear theory is a drastic approximation to physical
$QCD(3+1)$, the phenomenology of its DLCQ solutions demonstrate general
gauge theory features, such as the peaking of the wavefunctions at
minimal invariant mass, color coherence and the helicity retention of
leading partons in the polarized structure functions at $x\rightarrow
1$.

\section{Applications of Light-Cone Methods to QCD Phenomenology}

{\em Regge behavior.}\brk The light-cone wavefunctions $\psi_{n/H}$ of a
hadron are not independent of each other, but rather are coupled via
the equations of motion.  Recently Antonuccio, Dalley and I \cite{ABD}
have used the constraint of finite ``mechanical'' kinetic energy to
derive``ladder relations" which interrelate the light-cone
wavefunctions of states differing by 1 or 2 gluons.  We then use these
relations to derive the Regge behavior of both the polarized and
unpolarized structure functions at $x \rightarrow 0$, extending
Mueller's derivation of the BFKL hard QCD pomeron from the properties
of heavy quarkonium light-cone wavefunctions at large $N_C$ QCD
\cite{Mueller}.

{\em High momentum transfer exclusive reactions.} \brk Given the solution
for the hadronic wavefunctions $\psi^{(\Lambda)}_n$ with $\M^2_n <
\Lambda^2$, one can construct the wavefunction in the hard regime with
$\M^2_n > \Lambda^2$ using projection operator techniques \cite{BL}.
The construction can be done perturbatively in QCD since only high
invariant mass, far off-shell matrix elements are involved.  One can
use this method to derive the physical properties of the LC
wavefunctions and their matrix elements at high invariant mass.  Since
$\M^2_n = \sum^n_{i=1} \left(\frac{k^2_\perp+m^2}{x}\right)_i $, this
method also allows the derivation of the asymptotic behavior of
light-cone wavefunctions at large $k_\perp$, which in turn leads to
predictions for the fall-off of form factors and other exclusive matrix
elements at large momentum transfer, such as the quark counting rules
for predicting the nominal power-law fall-off of two-body scattering
amplitudes at fixed $\theta_{cm}.$ The phenomenological successes of
these rules can be understood within QCD if the coupling $\alpha_V(Q)$
freezes in a range of relatively small momentum transfer \cite{BJPR}.

{\em Analysis of diffractive vector meson photoproduction.} \brk The
light-cone Fock wavefunction representation of hadronic amplitudes
allows a simple eikonal analysis of diffractive high energy processes,
such as $\gamma^*(Q^2) p \to \rho p$, in terms of the virtual photon
and the vector meson Fock state light-cone wavefunctions convoluted
with the $g p \to g p$ near-forward matrix element \cite{BGMFS}. One can
easily show that only small transverse size $b_\perp \sim 1/Q$ of the
vector meson wavefunction is involved. The hadronic interactions are
minimal, and thus the $\gamma^*(Q^2) N \to \rho N$ reaction can occur
coherently throughout a nuclear target in reactions such as without
absorption or shadowing. The $\gamma^* A \to \phi A$ process thus
provides a natural framework for testing QCD color transparency
\cite{BM}.

{\em Structure functions at large $x_{bj}$.}\brk The behavior of structure
functions where one quark has the entire momentum requires the
knowledge of LC wavefunctions with $x \rightarrow 1$ for the struck
quark and $x \rightarrow 0$ for the spectators.  This is a highly
off-shell configuration, and thus one can rigorously derive
quark-counting and helicity-retention rules for the power-law behavior
of the polarized and unpolarized quark and gluon distributions in the
$x \rightarrow 1$ endpoint domain.  It is interesting to note that the
evolution of structure functions is minimal in this domain because the
struck quark is highly virtual as $x\rightarrow 1$; \ie\ the starting
point $Q^2_0$ for evolution cannot be held fixed, but must be larger
than a scale of order $(m^2 + k^2_\perp)/(1-x)$  \cite{BrodskyLepage}.

{\em Intrinsic gluon and heavy quarks.}\brk The main features of the heavy
sea quark-pair contributions of the Fock state expansion of light
hadrons can also be derived from perturbative QCD, since $\M^2_n$ grows
with $m^2_Q$.  One identifies two contributions to the heavy quark sea,
the ``extrinsic'' contributions which correspond to ordinary gluon
splitting, and the ``intrinsic" sea which is multi-connected via gluons
to the valence quarks. The intrinsic sea is thus sensitive to the
hadronic bound state structure \cite{IC}.  The maximal contribution of
the intrinsic heavy quark occurs at $x_Q \simeq {m_{\perp Q}/ \sum_i
m_\perp}$ where $m_\perp = \sqrt{m^2+k^2_\perp}$; \ie\ at large $x_Q$,
since this minimizes the invariant mass $\M^2_n$.  The measurements of
the charm structure function by the EMC experiment are consistent with
intrinsic charm at large $x$ in the nucleon with a probability of order
$0.6 \pm 0.3 \% $ \cite{HSV}.  Similarly, one can distinguish intrinsic
gluons which are associated with multi-quark interactions and extrinsic
gluon contributions associated with quark substructure \cite{BS}.  One
can also use this framework to isolate the physics of the anomaly
contribution to the Ellis-Jaffe sum rule.

{\em Rearrangement mechanism in heavy quarkonium decay.}\brk It is usually
taken for granted that a heavy quarkonium state such as the $J/\psi$
decays to light hadrons via the annihilation of the heavy quark
constituents to gluons. However, as Karliner and I \cite{BK} have
recently shown, the transition $J/\psi \to \rho \pi$ can also occur by
the rearrangement of the $c \bar c$ from the $J/\psi$ into the $\vert q
\bar q c \bar c >$ intrinsic charm Fock state of the $\rho$ or $\pi$.
On the other hand, the overlap rearrangement integral in the decay
$\psi^\prime \to \rho \pi$ will be suppressed since the intrinsic charm
Fock state radial wavefunction of the light hadrons will evidently not
have nodes in its radial wavefunction. This observation provides a
natural explanation of the long-standing puzzle why the $J/\psi$ decays
prominently to two-body pseudoscalar-vector final states, whereas the
$\psi^\prime$ does not.

{\em Asymmetry of Intrinsic heavy quark sea.}\brk As Ma and I have noted
\cite{BMa}, the higher Fock state of the proton $\vert u u d s \bar s>$
should resemble a $\vert K \Lambda>$ intermediate state, since this
minimizes its invariant mass $\M$.  In such a state, the strange quark
has a higher mean momentum fraction $x$ than the $\bar s$.
\cite{Signal,BMa} Similarly, the helicity intrinsic strange quark in
this configuration will be anti-aligned with the helicity of the
nucleon \cite{BMa}.  This $Q \leftrightarrow \bar Q$ asymmetry is a
remarkable, striking feature of the intrinsic heavy-quark sea.

{\em Direct measurement of the light-cone valence wavefunction.}\brk
Diffractive multi-jet production in heavy nuclei provides a novel way
to measure the shape of the LC Fock state wavefunctions. For example,
consider the reaction \cite{Bertsch,MillerFrankfurtStrikman}
\begin{equation}
\pi A \rightarrow {\rm Jet}_1 + {\rm Jet}_2 + A^\prime
\label{eq:h}
\end{equation}
at high energy where the nucleus $A^\prime$ is left intact in its
ground state.  The transverse momenta of the jets have to balance so
that $ \vec k_{\perp i} + \vec k_{\perp 2} = \vec q_\perp < \R^{-1}_A \
, $ and the light-cone longitudinal momentum fractions have to add to
$x_1+x_2 \sim 1$ so that $\Delta p_L < \R^{-1}_A$.  The process can
then occur coherently in the nucleus.  Because of color transparency;
\ie\ the cancellation of color interactions in a small-size
color-singlet hadron; the valence wavefunction of the pion with small
impact separation will penetrate the nucleus with minimal interactions,
diffracting into jet pairs \cite{Bertsch}.  The $x_1=x$, $x_2=1-x$
dependence of the di-jet distributions will thus reflect the shape of
the pion distribution amplitude; the $\vec k_{\perp 1}- \vec k_{\perp
2}$ relative transverse momenta of the jets also gives key information
on the underlying shape of the valence pion wavefunction.  The QCD
analysis can be confirmed by the observation that the diffractive
nuclear amplitude extrapolated to $t = 0$ is linear in nuclear number
$A$, as predicted by QCD color transparency.  The integrated
diffractive rate should scale as $A^2/\R^2_A \sim A^{4/3}$. A
diffractive experiment of this type is now in progress at Fermilab
using 500 GeV incident pions on nuclear targets \cite{E791}.

Data from CLEO for the $\gamma \gamma^* \rightarrow \pi^0$ transition
form factor favor a form for the pion distribution amplitude close to
the asymptotic solution \cite{BL} $\phi^{\rm asympt}_\pi (x) = \sqrt 3
f_\pi x(1-x)$ to the perturbative QCD evolution equation 
\cite{Kroll,Rad,BJPR} It will be interesting to see if the 
diffractive pion to di-jet experiment also favors the asymptotic form.

It would also be interesting to study diffractive tri-jet production
using proton beams $ p A \rightarrow {\rm Jet}_1 + {\rm Jet}_2 + {\rm
Jet}_3 + A^\prime $ to determine the fundamental shape of the 3-quark
structure of the valence light-cone wavefunction of the nucleon at
small transverse separation. Conversely, one can use incident real and
virtual photons: $ \gamma^* A \rightarrow {\rm Jet}_1 + {\rm Jet}_2 +
A^\prime $ to confirm the shape of the calculable light-cone
wavefunction for transversely-polarized and longitudinally-polarized
virtual photons.  Such experiments will open up a remarkable, direct
window on the amplitude structure of hadrons at short distances.

\end{document}